**Title:** Hydrogen release at metal-oxide interfaces: A first principle study of hydrogenated Al/SiO$_2$ interfaces

**Author:** Jianqiu Huang[1], Eric Tea[1], Guanchen Li[1], and Celine Hin[1,2]

[1] Department of Mechanical Engineering, Virginia Tech, Goodwin Hall, 635 Prices Fork Road - MC 0238, Blacksburg, VA 24061, USA

[2] Department of Material Science and Engineering, Virginia Tech, Goodwin Hall, 635 Prices Fork Road- MC 0238, Blacksburg, VA 24061, USA

**Abstract**

The Anode Hydrogen Release (AHR) mechanism at interfaces is responsible for the generation of defects, that traps charge carriers and can induce dielectric breakdown in Metal-Oxide-Semiconductor Field Effect Transistors. The AHR has been extensively studied at Si/SiO$_2$ interfaces but its characteristics at metal-silica interfaces remain unclear. In this study, we performed Density Functional Theory (DFT) calculations to study the hydrogen release mechanism at the typical Al/SiO$_2$ metal-oxide interface. We found that interstitial hydrogen atoms can break interfacial Al-Si bonds, passivating a Si $sp^3$ orbital. Interstitial hydrogen atoms can also break interfacial Al-O bonds, or be adsorbed at the interface on aluminum, forming stable Al-H-Al bridges. We showed that hydrogenated O-H, Si-H and Al-H bonds at the Al/SiO$_2$ interfaces are polarized. The resulting bond dipole weakens the O-H and Si-H bonds, but strengthens the Al-H bond under the application of a positive bias at the metal gate. Our calculations indicate that Al-H bonds and O-H bonds are more important than Si-H bonds for the hydrogen release process.

# 1. Introduction:

The reliability of silica-based devices remains the main concern of the semiconductor industry. Silica constitutes the basis of low-k materials that have been used extensively for the metal-oxide interconnect structure of integrated circuits (IC) to minimize signal delays [1,2]. Dielectric breakdown triggered by the oxide degradation [3-9] is one of the major failure mechanisms for these devices. Three primary defect generation models have been built to describe the dielectric breakdown: (i) electron impact ionization [10], (ii) Anode Hole Injection (AHI) [11], and (iii) Anode Hydrogen Release (AHR) [12-14]. The first two mechanisms occur when the gate voltage is much higher than the electron potential barrier height at interfaces. However, for ultrathin dielectric films, the breakdown usually takes place for gate voltage lower than the potential barrier height. The third mechanism, AHR, is the predominant defect generation mechanism when the gate voltage is lower than the electron potential barrier height. It is also the key factor responsible for the stress-induced leakage current (SILC) [12], which is commonly assisted by the thermal-heating of electron-induced vibrational modes [13,14].

Hydrogen becomes necessary in producing a high-quality interface as it passivates dangling bonds [15,16]. So, the hydrogen concentration in a device is commonly very high, especially at the interface having more dangling bonds than in bulk region. In addition to the AHR, many studies reported that hydrogen impurities also form various complex defects in the silica network [17-20]. The complex defects generate electron/hole traps, that assist the leakage current, and degrades materials. Consequently, the physics of hydrogen release and reaction process at the interface and their effects on breakdown have been extensively studied over the past decades, with a special focus on the Si/SiO$_2$ interface [4,12-14,19-21].

The hydrogen concentration at metal-oxide interfaces is considerably high, typically around ~10$^{14}$ cm$^{-2}$ [23]. Besides, the metal-silica interface also plays an irreplaceable role in billions of MOS transistors [24]. However, only a few studies have been previously published reporting the significance of hydrogen at the metal-oxide interface. Lin *et. al.* [22] speculated that the aluminum gate-anode might release positive hydrogen ions that degrade silica and generate traps, but provided no evidence to support their hypothesis. Therefore, our goal is to investigate the hydrogen release at metal-oxide interfaces from a case study, namely Al/SiO$_2$ metal-silica interfaces, using first-principles calculations.

The paper is organized as follow: we present the details of our computational setups, models, and techniques used for the analysis of the interface in Section 2. We then show our calculation results in Section

3. Results analysis and discussions are presented in Section 4. Finally, we draw the conclusion in Section 5.

2. Methodology

The Al/SiO$_2$ metal-oxide system with interfaces connecting Al (111) to $\alpha$-quartz (001) have been studied. As shown in Figure 1, the Al/SiO$_2$ interface has 224 atoms and contains 7 Al layers and 10 SiO$_2$ layers. The supercell contains a ~20 Å vacuum layer that separates the periodic image to minimize the slab-slab interaction. Hydrogen atoms have been used to passivate the Al and SiO$_2$ surface dangling bonds to lock the surface charge, keeping it in a bulk-like environment. The lattice mismatch between Al and SiO$_2$ at the interface is less than 1% so that the interface stress can be neglected.

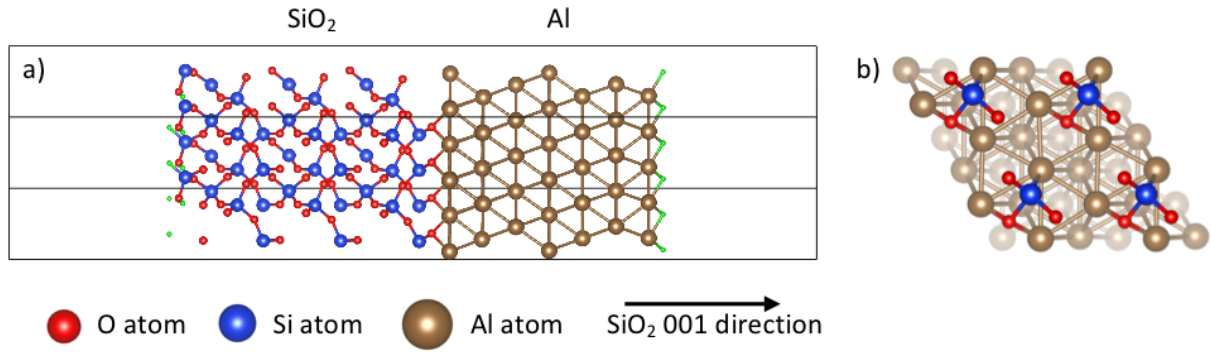

Figure 1 (color online): Sketches of the Al/SiO$_2$ interface model. Small (red), medium (blue) and large (brown) balls show O, Si, and Al atoms, respectively. a) ~10Å × 10Å × 55Å supercell containing 224 atoms with passivating H atoms shown by small balls at the surfaces (green). b) top view of the model showing the epitaxy of the interface in the SiO$_2$ (001) plane.

First principle calculations have been performed within the framework of Density Functional Theory (DFT) to study the interface model shown in Figure 1. The Projector Augmented Wave (PAW) method has been used, as implemented in the Vienna *Ab initio* Simulation Package (VASP) [25-27]. Our previous study showed that the Al/SiO$_2$ interface is metallic [28] and the Local Density Approximation (LDA) [29] exchange and correlation functional achieved a satisfactory energy and force minimization. So, the LDA have also been chosen to relax the hydrogenated Al/SiO$_2$ interface. The plane wave cutoff energy for our calculations is set to 400 eV, and the total energy convergence criteria to $10^{-5}$ eV. The conjugate gradient algorithm has been used to relax the atomic positions until forces on atoms are smaller than 10 meV/Å. The dipole-correction has been applied to compensate the macroscopic dipole generated by the asymmetry between two surfaces and the interface. This computational setup provides a reasonable convergence of the structures and total energies.

We analyze our DFT calculated charge density using Electron Localization Function (ELF) to investigate the bonding chemistry of hydrogen atoms at Al/SiO$_2$ interfaces. ELF measures the likelihood of finding another identical (same spin) electron in the vicinity of the reference electron, based on its kinetic energy density [30,31]. Thus, it can estimate the degree of localizations in a region of spaces having highly concentrated paired and unpaired electrons, which correspond to the bonding and nonbonding electrons (including dangling bonds and lone electron pairs). Visualizations of the volumetric data in this study are performed with VESTA [32].

We study the Bader charge of the system, which quantifies the bonding polarity, to explore the charge transfer of hydrogenated polar bonds at the interface. The Bader charge measures the quantity of charges within a region of spaces (Bader volumes) defined by interatomic surfaces having zero gradient vector of the spatial charge density [34]. So, the analysis of Bader charges provides reasonable information about the atomic charges, which allows us to quantitatively calculate the charge transfer of hydrogenated polar bonds at the interface. We calculate the Bader charge, using the code developed by the Henkelman's Group [35].

The AHR defect generation mechanism consists of two processes, (i) the release, and (ii) the reaction [13]. In the release process, injected electrons gain energy from electric fields and travel through oxide towards the anode, releasing protons (positively charge hydrogen, H$^+$) at the oxide/anode interfaces by losing their excess kinetic energy. In the reaction process, released H$^+$ ions diffuse backwards to the oxide/cathode interface (or are trapped in the oxide) and react with some precursors to generate defects (such as the hydrogen bridges), that will finally cause the breakdown. Therefore, the breakage of existing hydrogenated bonds in the release process is a principal concern for AHR induced breakdown.

In this study, we investigate the stability and bond strength of hydrogen at the Al/SiO$_2$ interfaces by studying the hydrogen impurity formation energy $E_f$ and the hydrogen cohesive (binding) energy $E_{coh}$. The hydrogen impurity formation energy, which measures the energy required to place a H atom at an interstitial site at interfaces, is given by [33]:

$$E_f = E_{IF}^H - (E_{IF} + \mu_H) \qquad (1)$$

where $E_{IF}^H$ and $E_{IF}$ are the total energies of the hydrogenated interface and the hydrogen-free interface, respectively. In Eq. (1), $\mu_H$ is the hydrogen chemical potential acts as an offset in our DFT study. In *ab*

*initio* studies, it is usually calculated at $T = 0$ K for a diluted gas, and is given by:

$$\mu_H = \frac{1}{2} E_{H_2} \qquad (2)$$

where $E_{H_2}$ is the energy of an isolated hydrogen molecule. The hydrogen cohesive energy, which measures the hydrogen bonding energy, is given by:

$$E_{coh} = E_{IF}^o + E_H - E_{IF}^H \qquad (3)$$

where $E_{IF}^o$ is the total energy of a hydrogenated interface without the H atom and $E_H$ is the energy of an isolated H atom. The cohesive energy we obtain in Eq. (3) is the energy required to break the hydrogenated bond and remove this H atom from the supercell, which is equivalent to the bond dissociation enthalpy at the absolute zero temperature.

The dielectric breakdown mechanism based on the bond breakage model proposed by McPherson and Mogul [9] states that the polar atomic structure of SiO$_2$ generates bond dipoles, which could respond to electric fields and alter the bond dissociation enthalpy. The bond dissociation enthalpy $\Delta H$ as a function of local electric fields $F_{loc}$ is given by [9]:

$$\Delta H = H_o - \Delta p \cdot F_{loc} \cdot \cos(\theta) - \frac{1}{2} \alpha F_{loc}^2 \qquad (4a)$$

$$F_{loc} = \frac{3 + \chi}{3} F_{ox} \qquad (4b)$$

where $H_o$ is the 0K dissociation enthalpy in absence of electric fields (equal to the $E_{coh}$), $\alpha$ is the electric polarizability, $\chi$ is the electric susceptibility, $F_{ox}$ is the applied electric field on the oxide layer, $\Delta p$ is the bond dipole, and $\theta$ is the angle between dipole moment and electric field. Thus, the bond dissociation enthalpy characterizes bond strengths under an applied electric field. It can be used to analyze the hydrogen release process. In Eq. (4a), the bond dipole $\Delta p$ is important to determine the bond dissociation enthalpy. The bond dipole is defined as the difference between the dipole generated by a bonding and a nonbonding pair of atoms. Thus, the bond dipole $\Delta p$ can be calculated from Bader charge by:

$$\Delta p = \sum (r_i - r_c) q_i \Big|_{bonded} - \sum (r_i - r_c) q_i \Big|_{unbonded} \qquad (5)$$

where $q_i$ and $r_i$ are the Bader charges and their center coordinates, and $r_c$ is the coordinate of the dipole center (conventionally set to the center of mass).

Using the impurity formation energy in Eq. (1) and the bond dissociation enthalpy in Eq. (4a), the hydrogen concentration $N_H$ and bond breakage probability $P_H$, considering a Boltzmann distribution are given by:

$$N_H = N_i \exp\left(-\frac{E_f}{k_B T}\right) \tag{6}$$

and

$$P_H = \exp\left(-\frac{\Delta H}{k_B T}\right) \tag{7}$$

where $N_i$ is the concentration of available sites for hydrogen atoms, $k_B$ is the Boltzmann constant, and $T$ is the temperature in Kelvin. Consequently, the hydrogen release concentration $C_H$ without thermal heating is expressed as:

$$C_H = N_H \cdot P_H = N_i \exp\left[-\frac{\Phi_H - \Delta p \cdot F_{loc} \cdot \cos(\theta)}{k_B T}\right] \tag{8a}$$

where $\Phi_H$ can be defined as the releasing potential, which depends on the released species and their environmental chemistry. $\Phi_H$ is given by:

$$\Phi_H = (E_{IF}^o - E_{IF}) + (E_H - \mu_H) \tag{8b}$$

Therefore, the hydrogen release concentration $C_H$, without thermal assistance provides a reasonable estimation on the importance of its release when it occupies different possible sites of the lattice at the interface. The $C_H$ at a given temperature is a function of bond breakage barrier, local bond dipole, and electric field. The $(E_H - \mu_H)$ term in Eq. (8b) is an environmental dependent offset.

3. Results

In Section 3.1, we present results for hydrogen-free interface, including the bonding chemistry and electronic properties. In Section 3.2, we study how the presence of hydrogen at the interface affects the bonding chemistry and electronic properties of the materials by comparing hydrogenated interfaces to hydrogen-free interface. Finally, in Section 3.3, we focus on the hydrogen impurity formation energy and cohesive energy for different interstitial sites to estimate the importance of hydrogen release at hydrogenated Al/SiO$_2$ interfaces.

### 3.1 Hydrogen-free Al/SiO$_2$ interface

Figure 2 shows the spatial charge density and ELF for two sectional slices of the supercell. The slice planes are defined by the Al-Si-O and the Al-O-Al interfacial structures. The white dot line in each contour plot indicates the interface position, separating the metal (above the dot line) from the oxide (below the dot line). Within the metal, as expected, the electrons are uniformly distributed forming an electron gas. Within the oxide, the electrons are mainly located around the O atoms. An explicit ELF attractor with a value over 0.8 is found surrounding the O atom, indicating the ionic bonding character between silicon and oxygen (Figure 2b). In Figure 2a, we also observe an electron density gradient between the Si atom and the Al atom at the interface. The ELF in this region approaches to 1. It indicates that the Si atom at interfaces bonds with the Al atom, which is polarized as well. Indeed, as noticed from Figure 1, the Si atom at the interface is attached to three O atoms leaving one $sp^3$ state open, which is occupied by electrons from the metal part.

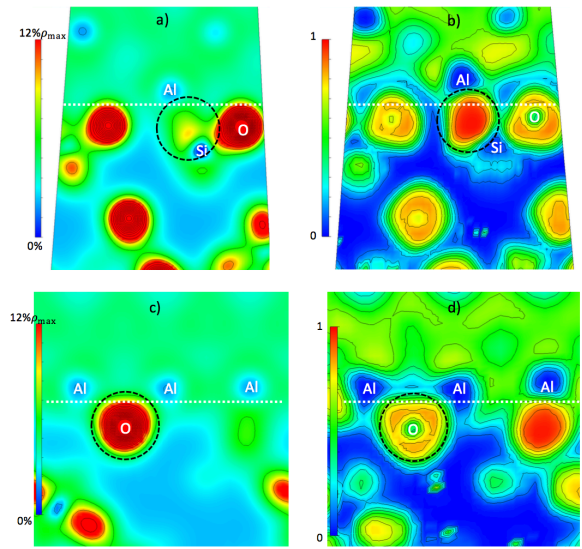

Figure 2 (color online): Contour plots of charge density (a) & (c) and ELF (b) & (d) for the hydrogen-free Al/SiO$_2$ interface. Plots a) and b) are taken in the interfacial Al-Si-O plane, and plots c) and d) are taken in the interfacial Al-O-Al plane. The white dot line in each plot indicates the interface position. Atoms intersected by the contour plane are labeled. Locally high charge concentration regions and explicit ELF attractors intersected by the contour plane are marked out by the black dot circle.

Figure 3 shows the Bader charges analysis, providing a quantitatively estimation on charge transfers. We assume that charge transfers only occur between the nearest neighbors. As shown in Figure 3, we conclude that:

- O atoms take away only 0.79 electrons from Si atoms, per Si-O bond at the interface,
- Each unit interface results in ~1.48 electrons transfer from aluminum to silica,
- Si atom at interfaces gains ~0.5 electrons from the Al atom at the left,
- Interfacial O atom gains 0.41 and 0.59 electron from the Al atom in the middle and at the right, respectively, forming two Al-O bonds.

These polarized interfacial bonds generate bond dipoles that would respond to external electric fields and affect the interface stability.

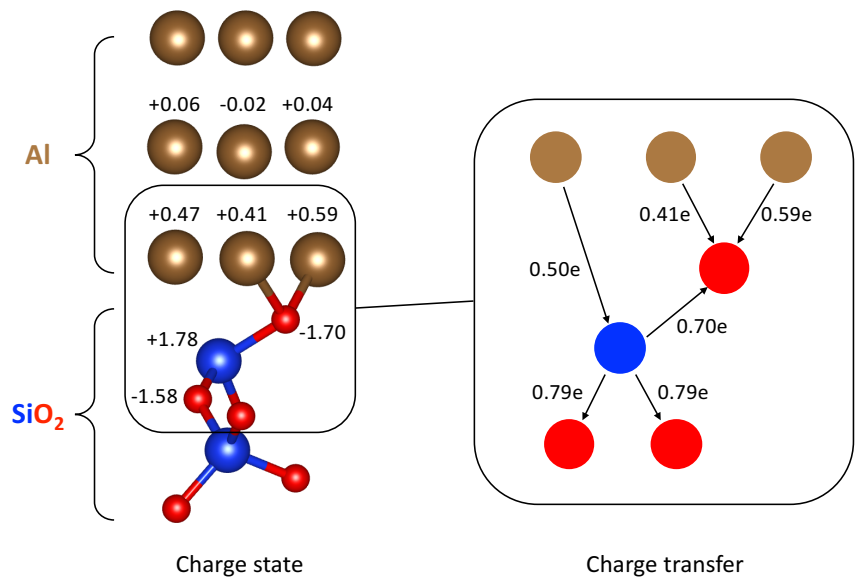

Figure 3 (color online): Charge states (left) and charge transfers (right) in a hydrogen-free interface unit: the left-hand side shows a cross-sectional view of the unit interface atomic charge state along (100) plane, and the right-hand side shows charge transfer of interface atoms. The charge state of Al, Si, and O atoms in bulks are 0, +3.14 and -1.57, respectively.

## 3.2 Hydrogen impurity, $I_H$

For the hydrogen-free Al/SiO$_2$ interface, we have shown that (i) silicon and oxygen bond with aluminum and (ii) the bonding at interfaces are polarized. In this section, we report the results for hydrogenated Al/SiO$_2$ interfaces. In Figure 4, we propose three hydrogenated configurations at the Al/SiO$_2$ interface, denoted as $I_H^{Al}$, $I_H^O$, and $I_H^{Si}$. These hydrogen interstitials are three potential sites where hydrogen can be adsorbed or released. It is worth noting that the charge transfer indicated along with the atomic configurations in Figure 4 represents the charge transfer relative to the hydrogen-free ones, revealing the transfers caused by the introduction of hydrogen impurities.

Figure 4a depicts the H atom that bridges two Al atoms to form the $I_H^{Al}$ interface, which leaves the remaining atoms nearly unperturbed. For this configuration, the Bader analysis shows that the aluminum atom transfers about 0.4 electrons to the hydrogen atom and form an Al-H bond. No other significant charge transfer is observed from the other neighbors. In addition, the charge density and ELF contours along the plane defined by the Al-H-Al structure in Figures 5a and 5b suggest that the Al-H is a polar bond. So, for the $I_H^{Al}$ interstitial case, hydrogen atoms at the interface form the polar bond with Al atoms. It should not affect the electronic property of Al/SiO$_2$ interfaces.

Figure 4b depicts the H atom that breaks one Al-O bond, and passivates the resulting O dangling bond to form the $I_H^O$ interface. This hydrogen atom likely interacts with a sub-interfacial O atom (denoted as sub-O), since the H atom attracts the sub-O atom to form a O-H-O structure. The Bader charge analysis indicates that the H atom transfers ~0.7 electrons to the O atom to form the polarized O-H bond. Besides, the O atom transfers ~0.57 electrons back to the Al atom for the Al-O bond breakage. There is zero charge transfer between the sub-O atom and the H atom. This suggests that the H atom and sub-O atom are not strongly bonded. This conclusion is emphasized by the high charge density around the H atom and the interfacial O atom, and the sparse charge density between the H and sub-O atoms, as shown in Figure 5c. In addition, in Figure 5d, the ELF contour also suggests that the H atom only bonds with the interfacial O atom. Thus, the unequal sharing of electrons gives the O-H polar bond a positive charge near the H atom, causing an electrostatic attraction between the H atom and the sub-O atom, which is known as a hydrogen bond.

Figure 4c depicts the H atom that breaks the Al-Si bond and passivates the Si dangling bond to form the $I_H^{Si}$ interface. No bond breakages or significant variation of atomic structure are observed for the surrounding atoms. The Bader charge analysis shows that the Si atom transfers ~0.58 electrons back to the

Al atoms, and ~0.67 electrons to the H atom. In addition, the charge density and ELF contours on the plane defined by the H-Si-O structure, as shown in Figures 5e and 5f, suggest $I_H^{Si}$ strongly localizes electrons around the Si atom. Therefore, when the hydrogen bonds to silicon at the interface, it locks the non-bonding silicon electron and closes the $sp^3$ orbital, forming polarized Si-H bonds.

Therefore, the H atom at metal-silica interfaces primarily passivates interfacial bonds or is adsorbed by the aluminum.

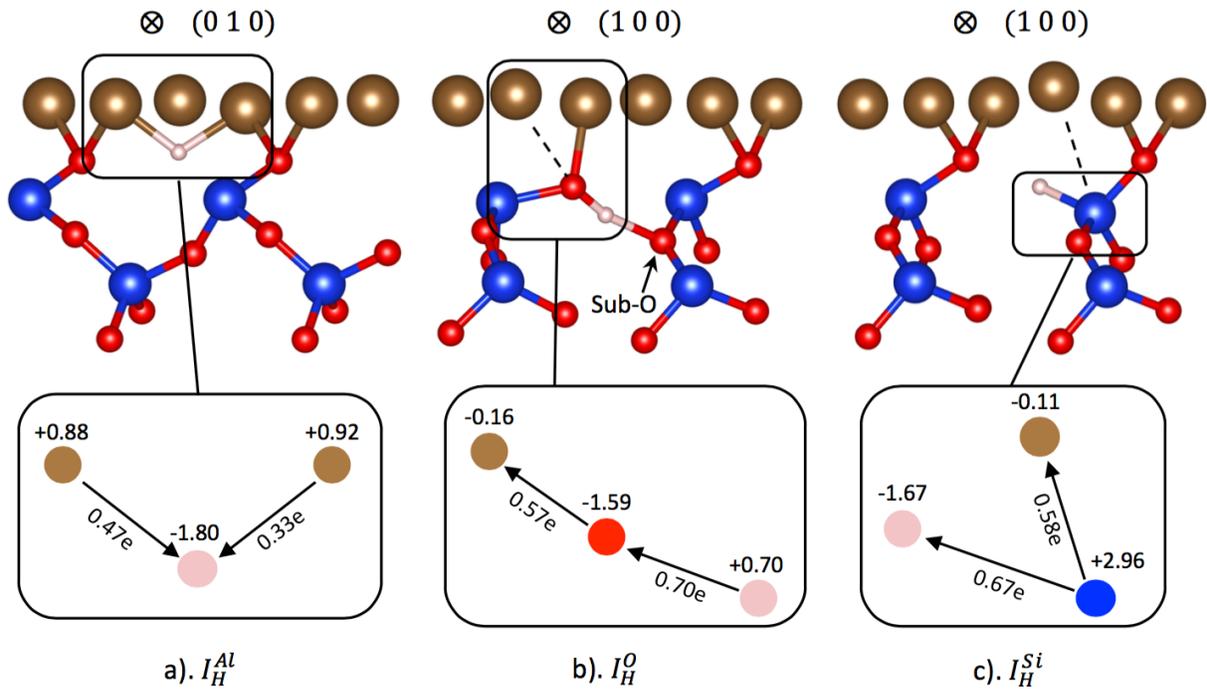

Figure 4 (color online): Cross-sectional views of the atomic structure and deduced charge transfers for $I_H^{Al}$, $I_H^O$, and $I_H^{Si}$ defective Al/SiO$_2$ interfaces. The plane Miller indexes are labeled at the top of each sketch with the direction of views pointing into the page. Major charge transfers are detailed in the inset below each sketch with arrows showing the transfer directions. Charge states are labeled above relevant atoms. a) The H atom bonds to two Al neighbors and does not cause bond breakage. b) The H atom breaks the original Al-O bond (the dash-line) and bonds to the O atom. Meanwhile, it forms a hydrogen bond with the second closest oxygen neighbor. c) The H atom breaks the original Al-Si bond (the dash-line) and bonds to the Si atom. For clarity purpose, Al-Si bonds are now displayed.

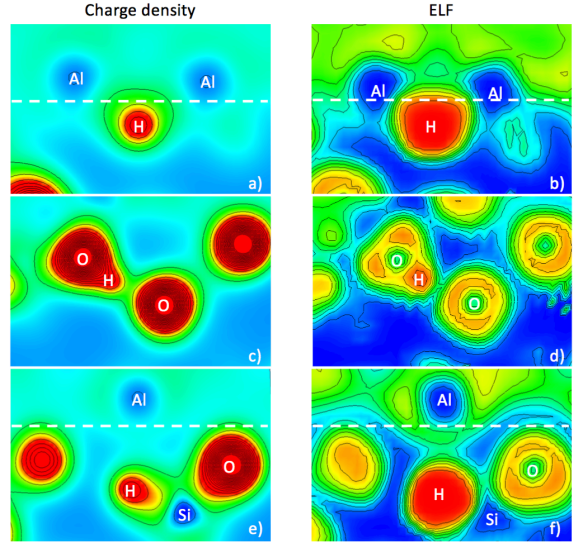

Figure 5 (color online): Contour plots of charge density (left) and ELF (right) for defective Al/SiO$_2$ interfaces. The white dashed line indicates the interface separating the metal (above the line) from the oxide (below the line). Plots a) and b) are taken in the Al-H-Al plane for $I_H^{Al}$. Plots c) and d) are taken in the O-H-O plane for $I_H^O$. Plots e) and f) are taken in the H-Si-O plan for $I_H^{Si}$. Atoms intersecting the contour planes are labeled.

### 3.3 Hydrogen stability and release

The strength and stability of hydrogenated bonds at the Al/SiO$_2$ interface dictate the hydrogen release concentration calculated with Eq. (8a). In Table I, we report the hydrogen impurity formation energy, $E_f$, and the hydrogen cohesive energy, $E_{coh}$. Calculations of $E_f$ using Eq. (1) show that hydrogen interstitials are more likely to form at the $I_H^{Si}$ site because of the lower formation energy. It is linked to the passivation of the non-bonding silicon -$sp^3$ orbital. According to Eq. (6), $I_H^{Si}$ sites have the highest concentration of H at the Al/SiO$_2$ interface. However, calculations of the hydrogen cohesive energy using Eq. (3) indicate that Al-H bonds have the lowest bond dissociation enthalpy, ~1.30 eV per Al-H bond. Therefore, $I_H^{Al}$ may be the interstitial defect that releases hydrogen the most easily. In order to obtain a picture that explicitly determines the primary source of hydrogen release, we analyze the hydrogen release potential $\Phi_H$ that comprises both factors of the hydrogen concentration and the bond dissociation enthalpy.

|  | $I_H^{Al}$ | $I_H^O$ | $I_H^{Si}$ |
|---|---|---|---|
| $E_f$ (eV) | -0.03 | -0.13 | -0.26 |
| $E_{coh}$ (eV) | 2.60 | 3.21 | 3.72 |

Table I: Hydrogen impurity formation energy $E_f$ and the hydrogen cohesive energy $E_{coh}$ at three different hydrogenated interfaces.

In order to determine the hydrogen release potential $\Phi_H$, we first need to calculate the bond dipole $\Delta p$. According to the bond breakage model proposed by McPherson and Mogul [9], bond dipole due to the polarity would respond to the external electric field and modify the bond dissociation enthalpy. Considering the nearest neighbor only, the bond dipole between two atoms can be calculated using Eq. (5). For each Al-H polar bond, the bond dipole moment has a value of ~0.82 Debye. The O-H bond generates a dipole moment to ~6.52 Debye. Although the hydrogen bond generates an electrostatic attraction between the H atom and the sub-O atom, as shown in Figure 4b, there is no net charge transfer between them resulting in a zero bond dipole. The Si-H bond also generates a dipole moment with a magnitude of ~0.83 Debye. Moreover, as shown in Figure 6, we can see that only Al-H bond dipoles point towards the metal, while the O-H and Si-O bond dipoles point towards the oxide. This suggests that the positive bias at the metal-gate increases the Al-H bond breakage enthalpy but reduce the bond breakage enthalpy of O-H and Si-O.

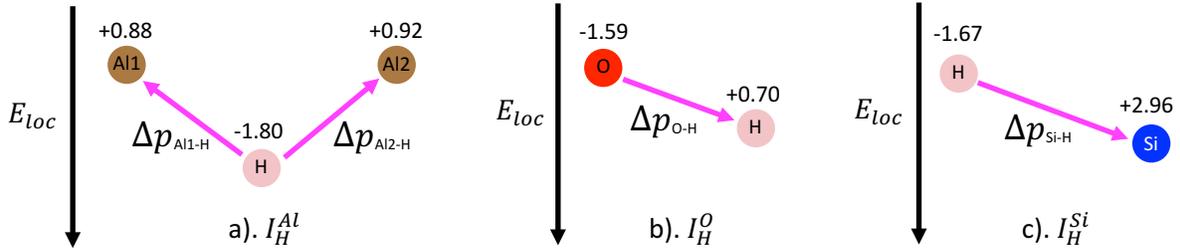

Figure 6 (color online): Bond dipole (pink arrows) generated by the polar hydrogenated bond at Al/SiO$_2$ interfaces. With the electric field (black arrows) from aluminum anodes, a) two Al-H dipoles resist the bond breakage, b) the O-H dipole reduces bond dissociation enthalpy, and c) the Si-H bond dipole also assists the bond breakage.

When we obtain the bond dipole, the hydrogen release potential $\Phi_H$ can be calculated using Eq. (8a). At the Al/SiO$_2$ interface, there are four interstitial sites available to form $I_H^{Al}$, $I_H^O$, and $I_H^{Si}$ impurities per supercell, setting $N_i$ in Eq. (8a) to 4. Therefore, the hydrogen release concentration $C_H$ from the different sites depends on exponent in Eq. (8a) only. We dropped the quadratic term of dipole moments in the calculation due to the extremely small electric polarizability of silica [9]. In Figure 7, we plot $\Phi_H$ as a function of $F_{loc}$ for $I_H^{Al}$, $I_H^O$, and $I_H^{Si}$ using the bond dipoles calculated above. Within the silica breakdown limit, approximately 10 MV/cm [4], the Si-H bonds appear to have the highest hydrogen release potential, and has a negligible response to electric fields. Meanwhile, the O-H bond has a ~0.5 eV lower release potential than the Si-H bond, but it is not the major source for hydrogen release. The major source for hydrogen release is the Al-H-Al bridge at the interface. It has the lowest release potential, even though its bond dipole strengthens Al-H bonds.

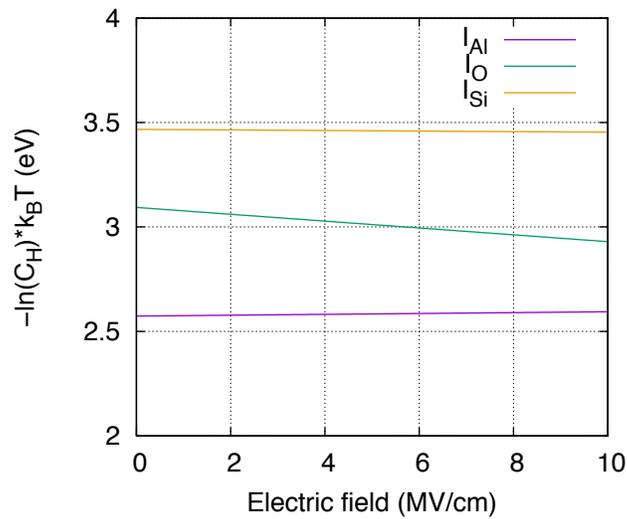

Figure 7 (color-online): Hydrogen release potentials as a function of the applied electric fields at room temperature for three type of Al/SiO$_2$ hydrogenated interfaces. One-half of the hydrogen molecule energy has been used for hydrogen chemical potential $\mu_H$ in the offset term $(E_H - \mu_H)$.

4. Discussion

The study on the hydrogen-free Al/SiO$_2$ interface reveals that aluminum transfers electrons into the silicon non-bonding -$sp^3$ orbital to form Al-Si bonds. These Al-Si bonds are not energetically stable, because they easily break in presence of hydrogen atoms at the interface. The hydrogen atoms passivate the Si non-bonding -$sp^3$ orbital and forms the Si-H bond, which results in the lowest total energy of the hydrogenated system. It suggests that hydrogen impurities are most stable at $I_H^{Si}$ sites. Thus, the Al-Si bond at the interface would be the major hydrogen trap, causing the $I_H^{Si}$ site to have the highest concentration of H.

Al-O bonds at the interface seem to be another hydrogen trap, because the O-H bond also lowers significantly the total energy of the system. So, the hydrogen concentration at $I_H^O$ sites may also be considerable at metal-silica interfaces. Our calculations, in Table I, have also shown that the hydrogen impurity at $I_H^{Al}$ sites has nearly zero formation energy. Also, previous experimental work showed evidences that the aluminum (111) surface adsorbs and stores hydrogen [36], so that $I_H^{Al}$ sites at the interface may have a significant hydrogen concentration as well. Therefore, the hydrogen concentration at $I_H^{Al}$, $I_H^O$ and $I_H^{Si}$ sites of the hydrogenated Al/SiO$_2$ interface is notably high. $I_H^{Si}$ sites have the highest hydrogen concentration and $I_H^{Al}$ sites have the lowest hydrogen concentration.

Even though the $I_H^{Si}$ site has the highest hydrogen concentration at metal-silica interfaces, it may not be the primary source of hydrogen release. The hydrogen release mechanism is not only related to the hydrogen concentration, but it also depends on the bond dissociation enthalpy. Calculations of the release potential indicates that Si-H bonds have the highest breakage potential resulting in the lowest possibility to release hydrogen atoms. On the contrary, the Al-H-Al bridge has the lowest hydrogen release potential, although the positive gate bias strengthens Al-H bonds. So, at the hydrogenated Al/SiO$_2$ interface, the $I_H^{Al}$ dominates the hydrogen release mechanism. However, different manufacturing process also affects the hydrogen concentration at the interface. Atomically growing aluminum on top of a desorbed silica surface would not allow hydrogen to adsorb at the aluminum surface. This would make the hydrogen concentration at $I_H^{Al}$ sites insignificant. With nearly zero hydrogen at $I_H^{Al}$ sites, the $I_H^O$ site becomes the most probable source of hydrogen release. Therefore, unlike at the Si/SiO$_2$ interface, the $I_H^{Si}$ is not the primary source of hydrogen release at the Al/SiO$_2$ interface.

The hydrogen atom at Al/SiO$_2$ interfaces breaks inhomogeneous Al-oxide bonds and locks non-bonding charges, stabilizing the system. However, once hydrogen atoms are released from the metal-silica interface, it starts to release H$^+$ ions that can diffuse through oxides towards the cathode interface. During the diffusion, the positive hydrogen ion could interact with intrinsic defects, such as oxygen vacancies, and generate a positively charged electron trap. A positively charged trap could enhance the local electric field and accelerate hydrogen release and diffusion. In addition, the increase in electron trap concentration in the oxide will also raise the SILC. Under such a scenario, the high electron current will increase the incoherent thermal heating [37] that also assists the hydrogen release as well. When hydrogenated bonds break at the Al/SiO$_2$ interface, all these sequential effects will continuously stress transistors until breakdown. Consequently, hydrogen atoms at the metal-silica interface play a key role for the degradation induced dielectric breakdown. The bonded hydrogen itself at interfaces does not significantly harm the system.

## 5. Conclusion

We performed DFT calculations on hydrogenated Al/SiO$_2$ interfaces to study the Anode Hydrogen Release mechanism. We found three energetically stable interstitial sites for hydrogen at the interface. We showed that polar bonds at the hydrogenated Al/SiO$_2$ interface creates bond dipoles, which alter the bond dissociation enthalpy. We conclude that Al-H bonds and O-H bonds are two major hydrogen sources for the hydrogen release mechanism. The Si-H bond is the least possible source for hydrogen release. Therefore, the introduction of hydrogen can produce high-quality interfaces, while it has to bear the risk of the degradation induced dielectric breakdown due to the Anode Hydrogen Release.

## Acknowledgement

This work was funded by the Air Force with program name: Aerospace Materials for Extreme Environment, and grant number: FA9550-14-1-0157. We acknowledge Advanced Research Computing at Virginia Tech for providing computational resources and technical support that have contributed to this work (http://www.arc.vt.edu).

## Reference

[1] K. Kapur, G. Chandra, J.P. McVittie, and K.C. Saraswat, "Technology and Reliability Constrained Future Copper Interconnects – Part IIL Performance Implications", *IEEE Transactions on Electron Devices*, 49 (2002) 598-604.

[2] M. Tada, N. Inoue, and Y. Hayashi, "Performance Modeling of Low-k/Cu Interconnects for 320-nm-Node and Beyond", *IEEE Transactions on Electron Devices*, 56 (2009) No. 9.

[3] J.W. McPherson, "Time dependent dielectric breakdown physics – Models revisited", *Microelectr. Reliab.*, 52 (2012) 1753-1760.

[4] S. Lombardo, J.H. Stathis, B.P. Linder, K.L. Pey, F. Palumbo, and C.H. Tung, "Dielectric breakdown mechanisms in gate oxides", *J. Appl. Phys.*, 98 (2005) 121301.

[5] C.-H. Ho, S.Y. Kim, and K. Roy, "Ultra-thin dielectric breakdown in devices and circuits: A brief review", *Microelectr. Reliab.*, 55 (2015) 308-317.

[6] S.P. Ogden, J. Borja, J.L. Piawsky, T.-M. Lu, K.B. Yeap, and W.N. Gill, "Charge transport model to predict intrinsic reliability for dielectric materials", *J. Appl. Phys.*, 118 (2015) 124102.

[7]   T. Usui, C.A. Donnelly, M. Logar, R. Sinclair, J. Schoonman, and F.B. Prinz, "Approaching the limits of dielectric breakdown for SiO$_2$ films deposited by plasma-enhanced atomic layer deposition", *Acta Materialia*, 61 (2013) 7660-7670.

[8]   F.-C. Chiu, "A Review on Conduction Mechanisms in Dielectric Films", *Advances in Materials Science and Engineering*, 2014 (2014) 578168.

[9]   J.W. McPherson and H.C. Mogul, "Disturbed Bonding States in SiO2 Thin-Films and Their Impact on Time-Dependent Dielectric Breakdown" IEEE IRPS, (1998) 47-56.

[10]  D.J. DiMaria, E. Cartier, and D. Arnold, "Impact Ionization, trap creation, degradation, and breakdown in silicon dioxide films on silicon", *J. Appl. Phys.*, 73 (1993) 3367.

[11]  D.J. DiMaria, E. Cartier, and D.A. Buchanan, "Anode hole injection and trapping in silicon dioxide", *J. Appl. Phys.*, 80 (1996) 304.

[12]  D.J. DiMaria and E. Cartier, "Mechanism for stress-induced leakage currents in thin silicon dioxide films", *J. Appl. Phys.*, 78 (1995) 3883.

[13]  E.Y. Wu and J. Sune, "Generalized hydrogen release-reaction model for the breakdown of modern gate dielectrics", *J. Appl. Phys.*, 114 (2013) 014103.

[14]  J. Sune and E.Y. Wu, "Hydrogen-Release Mechanisms in the Breakdown of Thin SiO2 films", *Phys. Rev. Lett.*, 92 (2004) 087601.

[15]  K.L. Brower, "Kinetics of H$_2$ passivation of $P_b$ centers at the (111) Si-SiO$_2$ interface", *Phys. Rev. B*, 38 (1988) 9657.

[16]  A. Constant, N. Camara, P. Godignon, and J. Camassel, "Benefit of H$_2$ surface pretreatment for 4$H$-SiC oxynitridation using N$_2$O and rapid thermal processing steps", *Appl. Phys. Lett.*, 94 (2009) 063508.

[17]  K. Doi, K. Nakamura, and A. Tachibana, "First-principle theoretical study on the electronic properties of SiO$_2$ models with hydrogenated impurities and charges", *Applied Surface Science*, 216 (2003) 463-470.

[18]  A. Yokozawa and Y. Miyamoto, "First-principles calculations for charged states of hydrogen atom in SiO$_2$", *Phys. Rev. B*, 55 (1997) 13783.

[19]  P.E. Blöchl and J.H. Stathis, "Hydrogen Electrochemistry and Stress-Induced Leakage Current in Silica", *Phys. Rev. Lett.*, 83 (1999 ) 372.

[20]  S.N. Rashkeev, D.M. Fleetwood, R.D. Schrimpf, and S.T. Pantelides, "Defect Generation by Hydrogen at the Si-SiO$_2$ interface", *Phys. Rev. Lett.*, 87 (2001) 165506.

[21]  S.N. Rashkeev, D.M. Fleetwood, R.D. Schrimpf, and S.T. Pantelides, "Dual behavior of H$^+$ at Si-SiO$_2$ interfaces: Mobility versus trapping", *Appl. Phys. Lett.*, 81 (2002) 1839.

[22]  C.-H. Lin, F. Yuan, B.-C. Hsu, and C.W. Liu, "Isotope effect of hydrogen release in metal/oxide/n-silicon tunneling diodes", *Solid-State Electronics*, 47 (2003) 1123-1126.

[23]  C. Kluthe, T. Al-Kassab, J. Barker, W. Pyckhout-Hintzen, and R. Kirchheim, "Segregation of hydrogen at internal Ag/MgO (metal/oxide)-interfaces as observed by small angle neutron scattering", *Act Materialia*, 52 (2004) 2701-2710.

[24]  G.E. Moore, "Gramming More Components onto Integrated Circuits", *Proceeding of the IEEE*, 86 (1998) 82-85.

[25]  P. E. Blöchl, "Projector augmented-wave method", *Phys. Rev. B*, 50 (1994) 17953.

[26]  G. Kresse and J. Furthmüller, "Efficiency of ab-initio total energy calculations for metals and semiconductors using a plane-wave basis set", *Comput. Mat. Sci.*, (1996) 6:15.

[27]  G. Kresse and J. Furthmüller, "Efficient iterative schemes for ab initio total-energy calculations using a plane-wave basis set", *Phys. Rev. B*, 54 (1996) 11169.

[28]  E. Tea, J. Huang, and C. Hin, "First principles study of band line up at defective metal-oxide interface: oxygen vacancy at Al/$SiO_2$ interface", *J. Phys. D*, 49 (2016) 095304.

[29]  J. P. Perdew and A. Zunger, "Self-interaction correction to density-functional approximations for many-electron systems", *Phys. Rev. B*, 23 (1981) 5048.

[30]  A.D. Becke and K.E. Edgecombe, "A simple measure of electron localization in atomic and molecular systems", *J. Chem. Phys.*, 92 (1990) 5397.

[31]  Y. Grin, A. Savin, and B. Silvi, "The ELF Perspective of chemical bonding", *The Chemical Bond: Fundamental Aspects of Chemical Bonding*, Ed. G. Frenking and S. Shaik, Weinheim: Wiley-VCH, 2014, pp. 345-382

[32]  K. Momma and F. Izumi, "VESTA 3 for three-dimensional visualization of crystal, volumetric and morphology data", *J. Appl. Crystallogr.*, 44 (2011) 1272-1276.

[33]  C. Freysoldt, B. Grabowsky, T. Hickel, and J. Neugebauer, "First-principles calculations for point defects in solids", *Reviews of Modern Physics*, 86 (2014) 253-305.

[34]  R.F.W. Bader, "Atoms in Molecules", *Acc. Chem. Res.*, 18 (1985) 9-15.

[35]  W. Tang, E. Sanville, and G. Henkelman, "A grid-based Bader analysis algorithm without lattice bias", *J. Phys.: Condens. Matter*, 21 (2009) 084204.

[36]  E.L. Crane and R.G. Nuzzo, "Collision-Induced Desorption and Reaction on Hydrogen-Covered Al(111) Single Crystals: Hydrogen in Aluminum?", *J. Phys. Chem. B*, 105 (2001) 3052-3061.

[37]  R.E. Walkup, D.M. Newns, and Ph. Avouris, "Role of multiple inelastic transitions in atom transfer with the scanning tunneling microscope", *Phys. Rev. B*, 48 (1993) 1858-1861.